\documentstyle[twoside,fleqn,espcrc2]{article}

\newcommand{\AmS}{{\protect\the\textfont2
  A\kern-.1667em\lower.5ex\hbox{M}\kern-.125emS}}

\title{Wave functions and their use in spectroscopy and phenomenology}

\author{Thomas A. DeGrand\address{Department of Physics,
        University of Colorado \\
        Boulder, Colorado 80309 USA}%
        and
        Matthew W. Hecht\address{National Center for Atmospheric Research \\
        P. O. Box 3000, Boulder, Colorado 80307 USA}}

\begin{document}

\begin{abstract}
We describe the calculation of Coulomb gauge wave functions
for light quark systems, and their use as interpolating fields
for excited state spectroscopy.
\end{abstract}

\maketitle

\section{WAVE FUNCTIONS}
We have been using wave functions reconstructed from lattice Monte Carlo
simulations of QCD to calculate the masses of orbital excitations.
We have also been trying to use wave functions as a tool for hadronic
phenomenology. Our work has been presented in a series of papers
\cite{ALLTDMH}. We present a synopsis of it here.
We will also comment on issues relevant to excited state spectroscopy which
came up during discussions during the conference.

\section{SPECTROSCOPY OF ORBITAL EXCITATIONS}

Lattice QCD is unique among the subfields of physics in focussing
only on  ground state spectroscopy.
 Some P-wave states' masses are regularly measured in
staggered simulations because they are the odd parity partners of
``ordinary'' states: the $a_1$ and $\rho$ are examples of such pairs.
In  nonrelativistic
QCD, Lepage and Thacker\cite{LEPAGET} have computed the masses of $\chi_C$
and $\chi_B$ states (without including spin effects).
The APE collaboration\cite{APE}  measured masses of
 some P-wave mesons in quenched simulations at $6/g^2 =\beta=5.7$, but
 had difficulty in continuing their program to higher
 $\beta$ \cite{APETWO}.
 Recently, the Fermilab group has presented a calculation
of the 1P-1S splitting in
charmonium, which they use to fix the strong coupling constant \cite{FNAL}.

We construct orbitally excited states by using
interpolating fields which couple only to a specific angular momentum
eigenstate, which are projected onto zero momentum and which
 are of large spatial extent to maximize overlap with the state.

At the $t \neq 0$ end
of the correlation function
we use an operator which depends on the relative separation of the
quarks, a ``wave function''\cite{ALLWF}.
The  wave function $\psi_G(r)$ of a hadron H in a gauge G
 is defined as
\begin{equation}
\psi_G(r) =
\sum_{\vec x} \langle H | q(\vec x)
{\bar q}(\vec x + \vec r) | 0 \rangle
\end{equation}
where $q(\vec x)$ and
${\bar q}(\vec y)$ are quantum mechanical operators
which create a quark and an antiquark at locations
$\vec x$ and $\vec y$. (We have suppressed Dirac and color indices.)
Our correlation function is  constructed from
 convolutions of quark and antiquark propagators
$G(x,y)$
\begin{eqnarray}
C(\vec r,t)  & = & \sum_{\vec x}\Psi(\vec y_1, \vec y_2)
G_q(\vec y_1,0;\vec x,t)\nonumber \\
 &  & G_{\bar q}(\vec y_2,0;\vec x + \vec r,t)
\label{eq:convol}
\end{eqnarray}
where $\Psi(\vec y_1, \vec y_2)$ is the $t=0$ operator.
At large $t$ if the mass of the hadron is $m_H$, then
\begin{equation}
C(\vec r,t ) \simeq \exp (-m_H t) \psi_G(\vec r)
\end{equation}
and so by plotting $C(\vec r, t)$ as a function of $\vec r$ we can reconstruct
the wave function up to an overall constant.
We measure the mass of a state by convoluting $C(\vec r,t)$ with some
  test function which further projects out the desired
state:
\begin{equation}
C(t) = \sum_{\vec r} \psi^*_{test}(\vec r) C(\vec r,t).
\label{eq:integ}
\end{equation}

At $t=0$ we take an operator which is separable in the coordinates
of the quarks. For a meson we use
\begin{equation}
\Psi(x_1,x_2) = \phi_1(x_1) \phi_2(x_2).
 \end{equation}
In order to couple to orbital excitations we take $\phi_1$
to be  an S-wave and $\phi_2$ to be some orbitally excited state
with angular momentum $l$,
 centered around some
specified coordinate.
 This state is a
linear superposition of a $\vec p = 0$ $L=l$ orbital excitation and a state
whose center of mass momentum is nonzero.
 (This is the familiar ``translation mode'' of a shell model state.)
Convoluting quark propagators as
in Eqn. \ref{eq:convol} removes the $\vec p \neq 0$ state and gives us
the wave function of the $\vec p = 0$ $L=l$ state.

Our trial states $\phi(x)$ were chosen to be Gaussians times an appropriate
spherical harmonic. We used Coulomb gauge.
We discussed with D. Richards whether P-wave spectroscopy could be done with
other sources and concluded that one could use Wuppertal sources
(see Ref. \cite{WUPP}): the S-wave $\phi$ could be the inverse of
$D^2 + m^2$ and the P-wave $\phi$ could be $\phi_P = D_i \Phi_S$.
This would eliminate any need to gauge fix or perform convolutions.
We have to say, however, that we prefer the possibility of reconstructing
the sink wave function at the end of the calculations; we have seen too
many QCD simulations where one discovered at the end of a lot of
data collection that one's
operators were not as effective at producing the desired state as one
would have wished.

There were several things that we should have done differently, which
should be part of a second generation simulation.  First, we only
recorded wave function information on a small number of time slices.
We took a trial function $\psi_{trial}(r)$ and recorded
the $C(t)$ resulting from Eqn. \ref{eq:integ}.
Had we possessed wave function information on all sites, we could have
reconstructed the wave function from the large $t$ $C(r,t)$ and used it
as the $\psi_{trial}(r)$ in Eqn. \ref{eq:integ}. This might have given a
better signal. Second, we used sources which were angular momentum eigenstates
only for the ${}^3P_2$, ${}^1P_1$, ${}^3D_3$, and ${}^1D_2$ mesons,
and the highest angular momentum baryons. The other states are merely
eigenstates of $m_J$.
This was done to limit the number of propagators constructed.
 We could have fully reconstructed the sink
angular momentum wave function, obtaining correlators which were eigenfunctions
of $j$ and $m_J$. We used nonrelativistic source and sink wave
functions (upper Dirac components in Bjorken-Drell basis). This is probably
good enough for heavier masses but might not be the optimal choice for
very light quarks. Finally, in excited state baryon spectroscopy
one must deal with the fact that many states with the same quantum numbers
are present in a multiplet. For example, there are two $j=1/2$ nucleons
in the $L=1$ [70] of SU(6). Separating these states presents a technical
challenge.

Our simulations used Wilson fermions on
$16^4$ lattices at a coupling $\beta=6.0$, in quenched approximation,
with 80 lattices for P waves and 50 for D waves.
They were performed at the Pittsburgh Supercomputing Center.

\begin{figure}[htb]
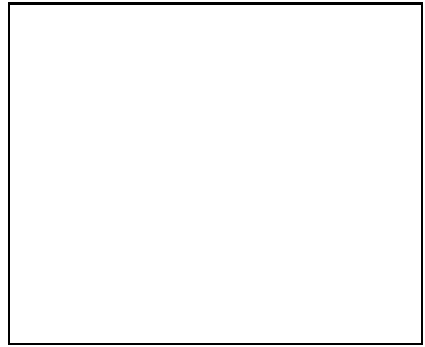

\framebox[55mm]{\rule[-21mm]{0mm}{43mm}}
\caption{Portrait of a ${}^3D_3$ meson.}
\label{fig:blah}
\end{figure}

We present a picture of a D-wave hadron in Fig. \ref{fig:blah};
 a slice through
a ${}^3D_3$ meson. The light regions are where the wave function
is large and positive; in the dark regions it is large and negative. The
characteristic quadrupole lobes are obvious. The orbitally excited
mesons and baryons are much larger than the S-wave states.
%

%

\begin{figure}[htb]
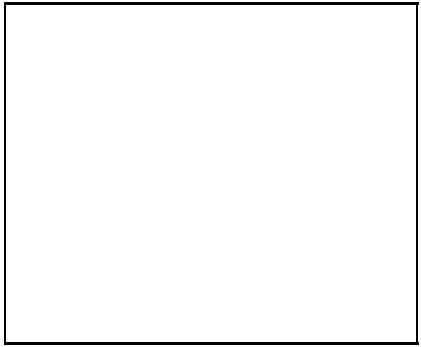

\framebox[55mm]{\rule[-21mm]{0mm}{43mm}}
\caption{Fine structure splitting at $\kappa=.1300$.}
\label{fig:finestructure}
\end{figure}

We were able to see a hint of P wave
fine structure splitting at our heaviest
quark mass. It is shown in Fig.  \ref{fig:finestructure}.
 With a nominal lattice spacing of
$1/a = 2$ GeV there is a remarkable similarity to charmonium, where the
${}^3P_0$ state is at 3415 MeV and the other states are at about 3500 MeV.

\begin{figure}[htb]
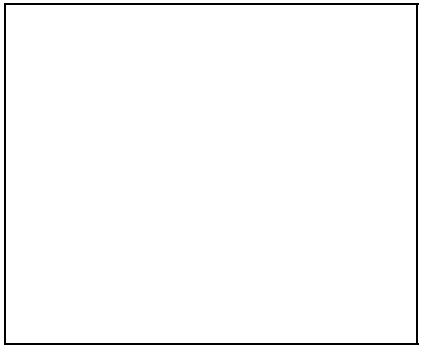

\framebox[55mm]{\rule[-21mm]{0mm}{43mm}}
\caption{Wilson spectroscopy vs. $\kappa$.}
\label{fig:walletcard}
\end{figure}

As a global way of presenting our results we show the ``Wilson fermion
wallet card'' in Fig. \ref{fig:walletcard}.
 We show S-wave mesons and baryons along with
the ${}^3P_2$, $N(5/2)$, ${}^3D_3$, and $N(7/2)$ excited states
(labelled ``p'', ``P'', ``d'', and ``D''.)

We extracted a lattice spacing by  lattice determinations
of the ${}^3P_2$ and ${}^3S_1$ states at $\kappa=0.130$, 0.145, and
0.152 and extrapolating their masses linearly in $\kappa$.
(We used the data of Ref. \cite{DL} for the $\kappa=.1300$ vector meson).
We determined the lattice spacing by fitting the extrapolated masses
to the  $\psi(3095)$ and $\chi(3555)$ masses.  This gave
a charm hopping parameter of $\kappa=.1224$ and an inverse lattice spacing
of $1/a=1790$ MeV. (Determining the lattice spacing from
our proton mass  would give $1/a=1710$ MeV; from the rho,
$1/a=2264$ MeV.)
The extrapolated common D-meson mass is then
3.99(16) MeV, where the error is only from the extrapolation.
The ${}^3D_1$ $c \bar c$ state is at 3.77 GeV but its mass
is influenced by the nearby $D \bar D$ threshold.
 Model calculations\cite{DWAVE}
 of D-wave states (some of which are narrow since their
decays to $D \bar D$ are forbidden) give masses of 3.81 to 3.84 GeV.
At this value of the lattice spacing our ${}^3P_2-{}^3P_0$ mass splitting
at $\kappa=.1300$ is 63 MeV; in charmonium the corresponding
number is 145 MeV.

This is a little different from the Fermilab procedure of extracting
a lattice spacing from the ${}^1P_1$ splitting from the center of gravity
of the S-wave mesons. We made our choice because we have a better signal
in the ${}^3P_2$ channel; anyway, the most robust experimental
 P-wave spectroscopy
pretty much across the board is in the  ${}^3P_2$ channel.

\section{PHENOMENOLOGY}
\begin{figure}[htb]
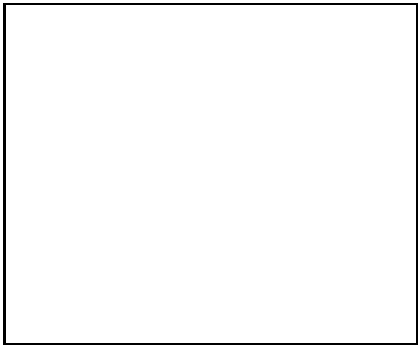

\framebox[55mm]{\rule[-21mm]{0mm}{43mm}}
\caption{Neutron/proton charge radius ratio.}
\label{fig:radius}
\end{figure}

While wave functions in a smooth gauge have clearly demonstrated their worth as
interpolating fields for spectroscopy, it is still not clear if they have
any real physical significance.  We compared charge radii computed using
wave functions with form factor data. The calculation is simple:
for mesons one computes
\begin{equation}
\langle r^2 \rangle = \sum_i e_i \int d^3 r |\psi(r)|^2  (r/2)^2
\label{eq:char}
\end{equation}
and there is an analogous equation for baryons. We found that
$\langle r^2 \rangle$ computed from Eqn. \ref{eq:char}
 was about a factor of four
to six smaller than the observed pion or nucleon charge radius.
However, at the same time, we saw that the ratio of neutron to
proton charge radii was negative and about the same size as the
experimental ratio. Apparently this negative neutron charge radius
arises because the two $d$ quarks have a larger relative separation than
the $u$ quark does from either of the $d$'s. This effect is easy
to explain as being due to the color hyperfine interaction (compare Ref.
\cite{ELLIS}).
We show an example of this ratio in Fig. \ref{fig:radius}.
\section{CONCLUSIONS}
The methods we describe here can and should be used for a large scale
calculation of the spectroscopy of orbital excitations.
We believe our signal for P-wave states is only limited by
statistics. However, the physical meaning
of the wave function remains obscure.

\section{ACKNOWLEDGEMENTS}
This work was supported by the U. S. Department of Energy. Simulations
were performed at the Pittsburgh Supercomputing Center. T. D. would like
to thank
A. El-Khadra,
D. Richards,
and
H. Thacker
for interesting conversations about wave functions during Lattice 92.

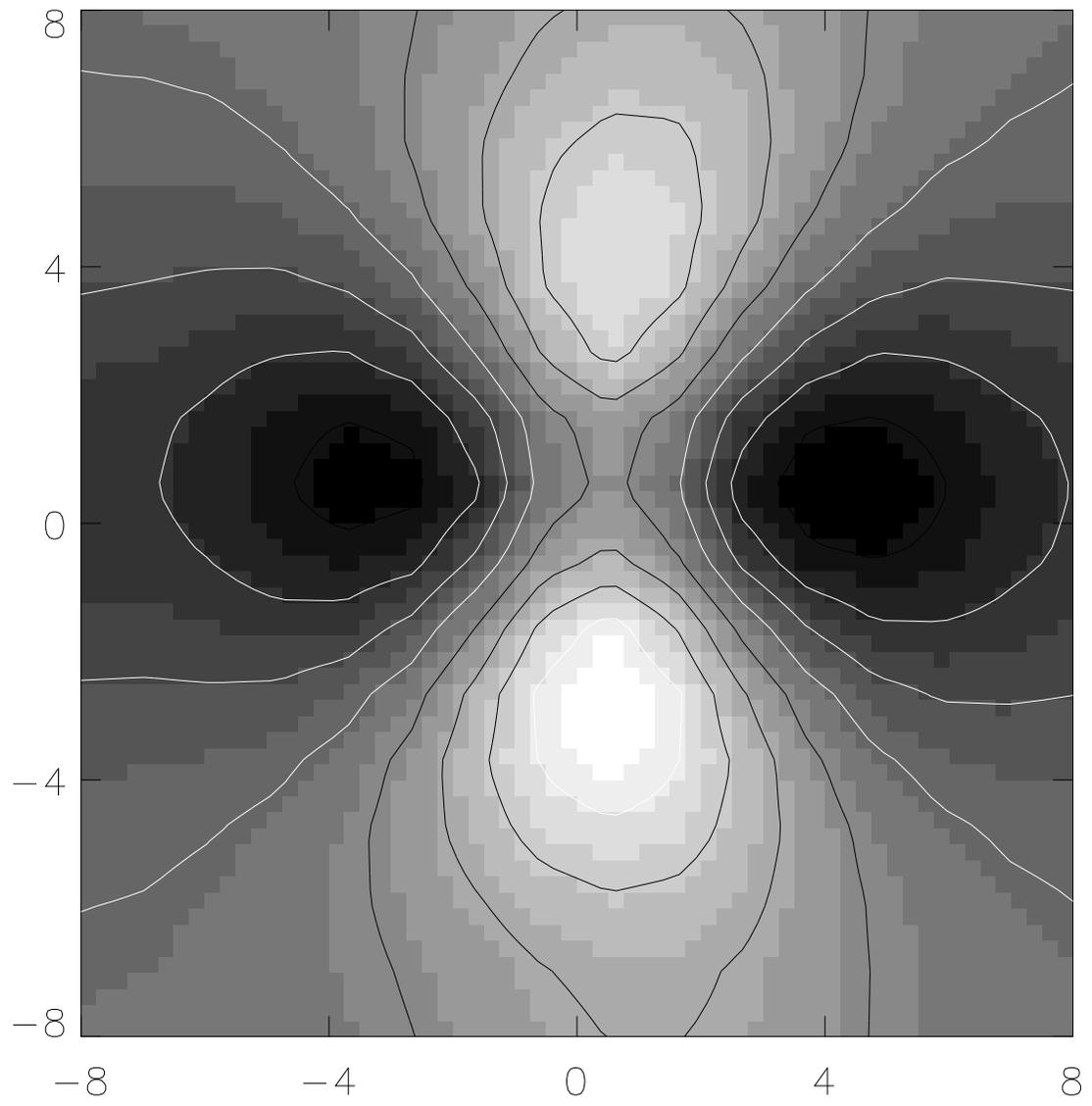
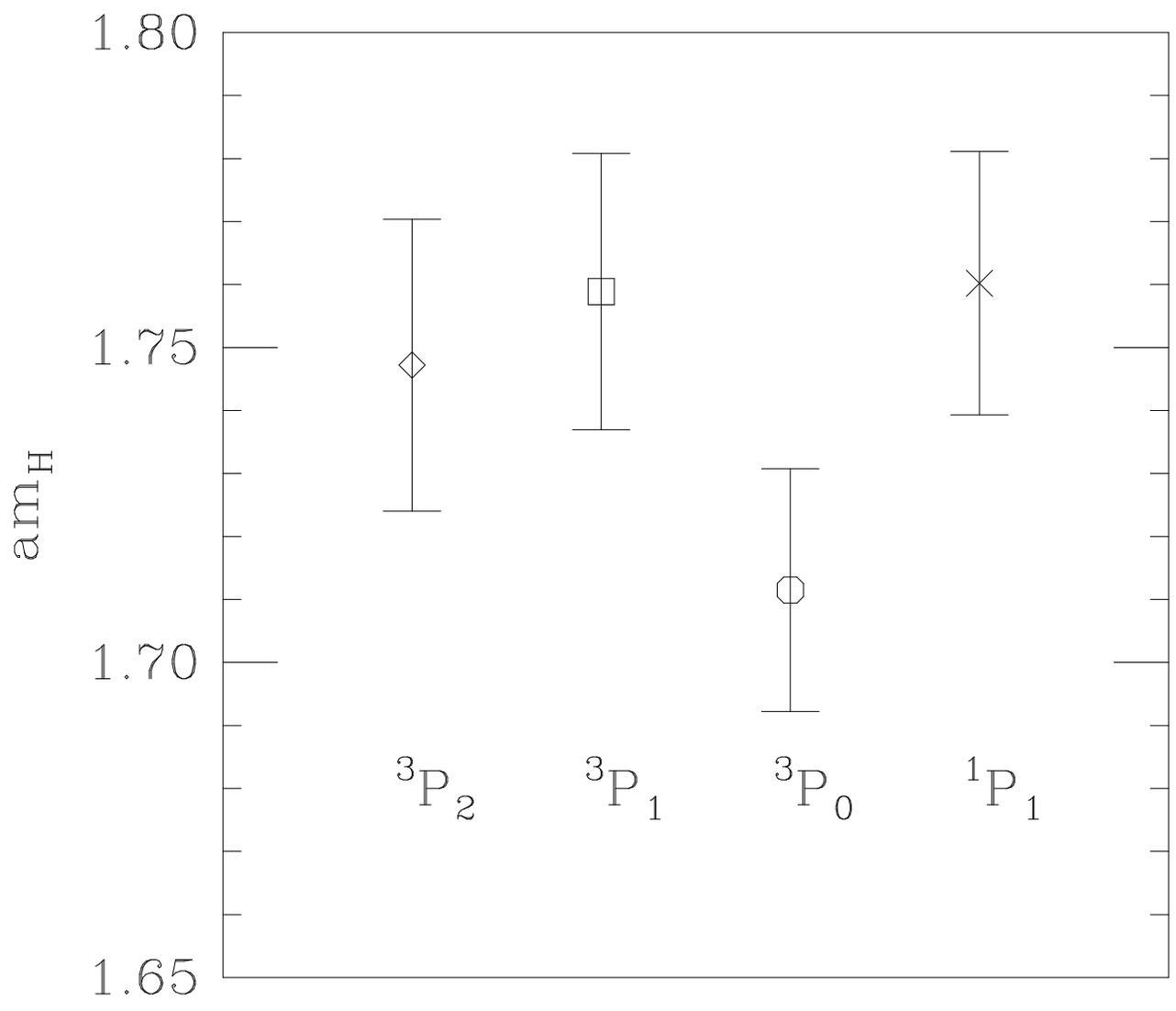
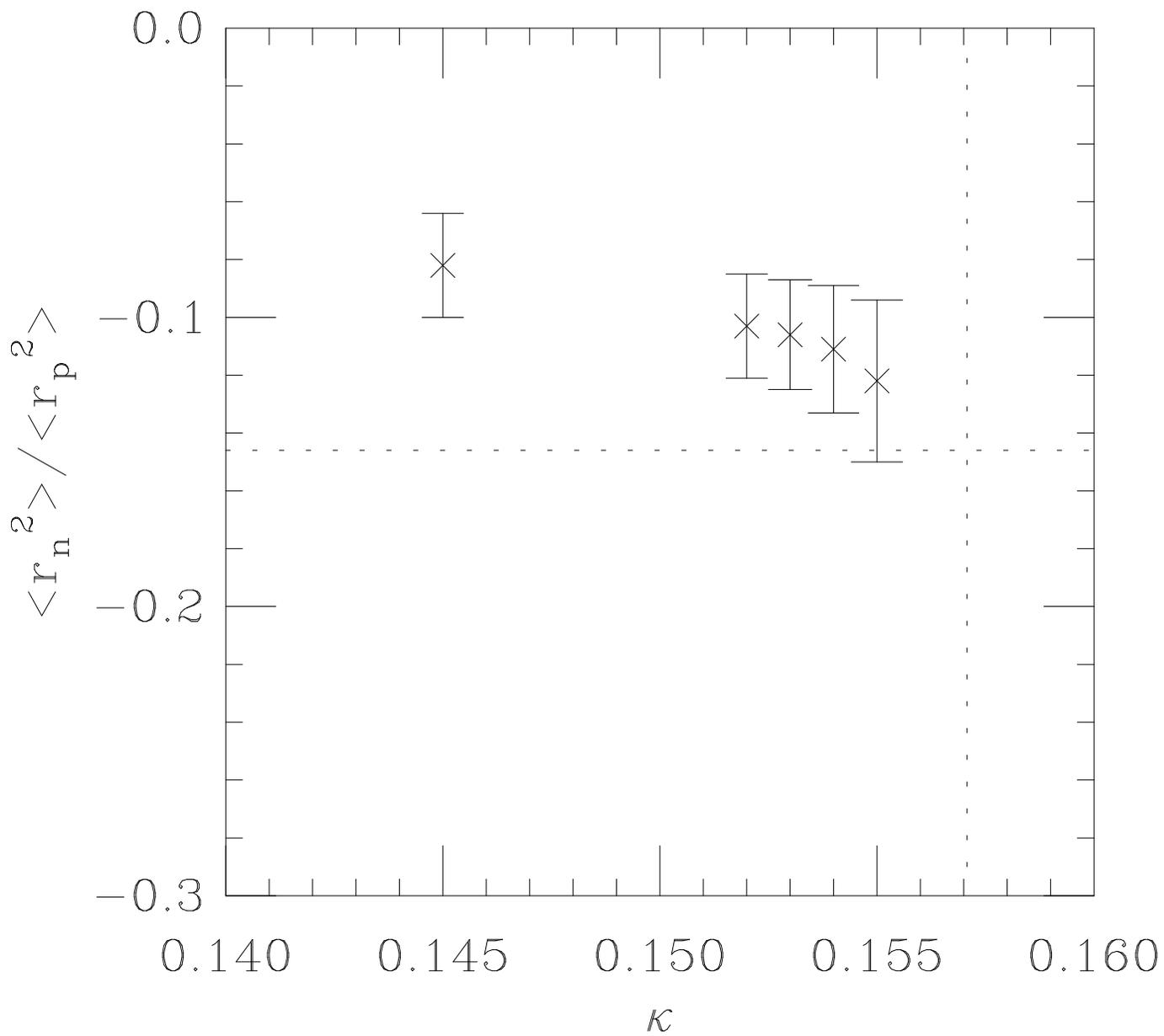
\end{document}